\documentclass[preprint,aps,showpacs]{revtex4}
\usepackage{amssymb}
\usepackage{epsfig,amsmath}
\usepackage{graphicx}
\usepackage{dcolumn}
\usepackage{bm}

\newcommand{\beq}{\begin{equation}}
\newcommand{\eeq}{\end{equation}}
\newcommand{\beqa}{\begin{eqnarray}}
\newcommand{\eeqa}{\end{eqnarray}} 
 
 \newcommand{\ka}{\kappa}

\newcommand{\la}{\langle}
\newcommand{\ra}{\rangle}

\def\ajp#1{{ Am.\ J.\ Phys.} {\bf #1}}

\def\nat#1{{ Nature} {\bf#1}}

\def\npho#1{{Nature\ Phot.} {\bf#1}}

\def\oc#1{{ Opt.\ Commun.} {\bf#1}}
\def\ol#1{{ Opt.\ Lett.} {\bf#1}}
\def\pla#1{{ Phys.\ Lett. A\/} {\bf#1}}

\def\pra#1{{ Phys.\ Rev. A\/} {\bf#1}}

\def\pre#1{{ Phys.\ Rev. E\/} {\bf#1}}
\def\prl#1{{ Phys.\ Rev.\ Lett.} {\bf#1}}

\def\rmp#1{{ Rev. \ Mod. \ Phys.} {\bf#1}}

\begin{document}

\title{Violation of Bell's Inequalities with Classical Shimony-Wolf
States: Theory and Experiment}
\author{Xiao-Feng Qian}
\author{Bethany Little, John C. Howell}
\author{J.H. Eberly}
\affiliation{Center for Coherence and Quantum Optics and the Department of
Physics \&
Astronomy\\
University of Rochester, Rochester, New York 14627}
\date{\today }

\begin{abstract}
For many decades the word ``entanglement" has been firmly attached to the
world of quantum mechanics, as is the phrase ``Bell violation". Here we
introduce Shimony-Wolf fields, entirely classical non-deterministic
states, as a basis for entanglement and Bell analyses. Such fields are
well known in coherence optics and are open to test. We present
experimental results showing that Shimony-Wolf states exhibit strong
classical Bell
violation, in effect opening a way of examining a new sector of the
boundary between quantum and classical physics.
\end{abstract}


\maketitle


The difference between classical and quantum phenomena is sometimes
difficult to pin down because distinctions can be prescribed in a number
of inequivalent ways. This is a reason for the notorious obscurity of the
quantum-classical border. State superposition is an essential quantum
attribute, but it is not exclusive to quantum physics since all linear
wave phenomena share it. Entanglement is often regarded as the
quintessentially non-classical aspect of the physical world, and a
quantum-classical distinction is provided by Bell-violation experiments.
Here we report a theoretical analysis and a related experiment regarding
the quantum-classical border as probed by Bell tests with classical waves.
We employ what we suggest can be named Shimony-Wolf fields or states for
this purpose.

First we note that observations, demonstrations and even applications of
non-quantum wave entanglement exist.  They exploit non-separable
correlations among two or more degrees of freedom (DOF) of optical wave
fields. In the past few years such applications have achieved notable
successes including the resolution of a long-standing open issue
concerning Mueller matrices \cite{Simon-etal}, unification of competing
interpretations of degree of polarization \cite{Qian-EberlyOL}, and
application of  the Bell measure as a new index of coherence in optics
\cite{Kagalwala-etal}. These developments followed even earlier
explorations of  non-separable DOF correlation, both theoretical and
experimental, in optical wave fields \cite{Spreeuw, Ghose-Samal01,
Lee-ThomasPRL, Borges-etal, Goldin-etal}.

Next we call attention to Shimony's reviews \cite{Shimony} of the
consequences of Bell's inequalities. He identifies three facts of quantum
Nature that must be recognized by any system $S$ produced for testing. In
his words, these are:\\
(I) In any state of a physical system $S$ there are some
eventualities which have indefinite truth values.\\ (II) If an
operation is performed which forces an eventuality with indefinite
truth value to achieve definiteness [...] the outcome is a matter of
chance.\\ (III) There are `entangled systems' (in Schr\"odinger's
phrase) which have the property that they constitute a composite
system in a pure state, while neither of them separately is in a pure
state.  (Here by eventualities Shimony means measurement outcomes.)

As it happens, within the well-known classical theory of optical coherence
(see Wolf \cite{Wolf59}) there are statistical states that satisfy all
three criteria. One quickly sees that the usual expression for the
classical electric field vector of a
transverse wave,
\beq \label{Efield}
\vec E = \hat x E_x(\vec r,t) + \hat y E_y(\vec r,t),
\eeq
is an entangled combination of the DOF for polarization and transverse
amplitude, and because the amplitudes are understood as stochastic
variables, the field takes a definite value only when it is observed.
Beyond its probabilistic indeterminacy, the $\vec E$ in (\ref{Efield})
shares other quantum attributes -- it has the form of a quantum state and
can be called a pure state in the same sense. It is really a bi-vector,
linear in two
vector spaces at once, lab space for $\hat x$ and $\hat y$, and
continuous function space for $E_x$ and $E_y$.  We will call it a
Shimony-Wolf field or state.

By using Shimony-Wolf states we are departing from the recently observed
applications of non-separable but also non-stochastic DOF correlations and
make a test of their stochastic extensions and limitations. The boundary
zone between classical and quantum physics is opened for examination in a
new way. We can address questions such as:  which features considered to
be intrinsic to quantum theory can be fully replicated in a classical
context? and  what is the role of Bell inequalities for Shimony-Wolf
states?

A prompt response to such questions could be to say that existing
observations of Bell Inequality violations \cite{Freedman-Clauser,
Aspect-etal, Ou-Mandel, Shih-Alley, Rowe-etal} argue that classical
systems are unable to provide the strong correlations predicted by quantum
theory and attained when tested.  The reply is that all such tests were
made by particle detection, which is not the subject here. It is known
that Bell inequalities can be tested with DOF-entangled classical wave
fields, as was demonstrated by Borges, et al. \cite{Borges-etal}, for
example. But such tests of classical fields have all employed a field
similar to $\vec E = \hat v\ \psi_v(\vec r) + \hat h\ \psi_h(\vec r)$,
where $\psi_v(\vec r)$ and $\psi_h(\vec r)$ are prescribed orthogonal
field modes.
Their determinate character, lack of statistical indefiniteness, means
that such fields can be written in fully separable form, $\vec E = \hat u
F(\vec r)$, at any location in the wave field -- it is fully factorable at
position $\vec r$ (the same as 100\% polarized) in the direction $\hat u$
defined by $\tan\theta = |\psi_v(\vec r)/\psi_h(\vec r)|$.


We will adopt Dirac-type notation for Shimony-Wolf vectors: $\vec E \to
|{\bf E}\ra$, $\hat x \to |x\ra$, etc., where we use boldface to emphasize
the two-space character of the field:
\beq
|{\bf E}\ra = |x\ra | E_x\ra + |y\ra |E_y\ra.
\eeq
We designate $I = \la {\bf E}| {\bf E}\ra = \la E_x| E_x\ra + \la E_y|
E_y\ra$ as the intensity. To treat any partially coherent optical field
(e.g., sunlight), we engage the powerful Schmidt Theorem \cite{Schmidt}
which allows us to write the intensity-normalized classical field as:
\beq
\label{e-def} |{\bf E}\ra/\sqrt{I}\ \equiv\ |{\bf e}\ra =
\kappa_{1}|u_1\ra|f_1\ra +
\kappa_{2}|u_2\ra|f_2\ra,
\eeq
where the real $\kappa_j$ satisfy $\kappa_1^2 + \kappa_2^2 = 1$. The
$|u\ra$s are Schmidt-rotated versions of $|x\ra$ and $|y\ra$ in lab space,
and the $|f\ra$s are linear superpositions of (typically infinitely many)
orthogonal functions making up the field components  \cite{Kac-Siegert}.
Significantly, there are only two $|f\ra$s that enter because the Schmidt
Theorem selects the only plane in the infinite-dimensional continuous
function space that is relevant for combination with $|x\ra$ and $|y\ra$.
In effect, an optimal two-way renormalization has been made, which yields
pairs of orthogonal unit vectors for both the polarization and amplitude
DOF:  $\la u_1 |u_2\ra = \la f_1 |f_2 \ra = 0$. The coefficients
$\kappa_{1}$ and $\kappa_{2}$ account for the different partial
intensities of the two terms, and they equal $1/\sqrt{2}$ in the
completely incoherent (fully unpolarized) case.

We now demonstrate that these classical Shimony-Wolf states have much more
than a notational relation with quantum theory and are ideally suited for
probing the specific quantum-classical interface defined by Bell
inequalities. Bell's agenda \cite{Bell64-66, BellSpeakable} led him to
focus on the
joint probabilities and correlations across two vector spaces, and
the Clauser-Horne-Shimony-Holt (CHSH) inequality \cite{CHSH} provides
the best-known mechanism for this.


The CHSH inequality arises from a combination of correlations defined
by a set of controllable parameters. In most cases these are angles
determined by detections in various rotated bases of the combined
vector spaces. For example, in the case of Shimony-Wolf state
(\ref{e-def}), arbitrary rotations of the Schmidt-chosen polarization
vectors $|u_1\ra$ and $|u_2\ra$ will be written
\beqa \label{uRotation}
&&|{u}_{1}^{a}\rangle =\cos a|{u}_{1}\rangle -\sin a|{u}_{2}\ra \quad
{\rm and} \notag \\
&& |{u}_{2}^{a}\ra =\sin a|{u}_{1}\ra +\cos a|{u}_{2}\ra,
\eeqa
where $a$ is the rotation angle. The two vectors obviously remain
orthogonal for any angle $a$: $\la {u}_{1}^{a}|{u}_{2}^{a}\ra = \la
{u}_{1}|{u}_{2}\ra = 0$. For function space rotations we have
$|f_{1}^{b}\ra$ and $|f_{2}^{b}\ra$ defined similarly, where the rotation
angles $a$ and $b$ are unrelated.

Finally, the joint correlation between the lab ($u$) and function
($f$) spaces can be written
\beq \label{CAB1}
C(a,b)=\la {\bf e}|Z^u(a)\otimes
Z^f(b)|{\bf e}\ra,
\eeq
where $Z^u(a) \equiv |{u}^{a}_1\ra\la{u}^{a}_1| -
|{u}^{a}_2\ra\la{u}^{a}_2|$ is the difference of two projectors
(analogous to the $z$ component of a $\vec \sigma$ operator).
$C(a,b)$ is thus a combination of various joint probabilities such as:
\beqa \label{Pklab}
P_{11}(a,b) &=& \la {\bf e}|\Big(|{u}_{1}^{a}\ra |f_{1}^{b}\ra \la
f_{1}^{b}|\la {u}_{1}^{a}|\Big)|{\bf e}\ra \\ \nonumber
&=& |\la f_{1}^{b}; {u}_{1}^{a}|{\bf e}\ra |^2.
\eeqa
The probabilities $P_{kl}(a,b)$ with $k,l=1,2$ all have familiar roles in
classical
statistical optics \cite{Brosseau-Wolf}.

Gisin \cite{Gisin} observed that any quantum state entangled in the same
way as the Shimony-Wolf pure state (\ref{e-def}) will permit CHSH
inequality violation. The same result is true here, as one uses only DOF
independence and properties of positive probabilities and normed vectors
to arrive at it.  We adopt the same approach \cite{SupplMat} and obtain
the familiar CHSH result:
\beqa \label{CHSH1}
&&-2\leq {\cal B} \leq 2, \quad {\rm where,\ as\ usual,} \nonumber \\
&&{\cal B} = C(a,b) - C(a,b') + C(a',b) + C(a',b'),
\eeqa
and $a,\ a',\ b,\ b'$ are arbitrary rotation angles.\
The only unfamiliar feature is that $\la {\bf e}|Z^u(a)|{\bf e}\ra $
and $\la {\bf e}|Z^f(b)|{\bf e}\ra$ both lie anywhere in the
continuum between $-1$ and $+1$, rather than taking the discrete
values $\pm 1$, since we have no quantum particles to be detected or
counted, but rather classical light beams with various intensities.


We now describe a sequence of Bell test experimental measurements with a
classical
non-deterministic Shimony-Wolf wave field. The experiment is
designed to evaluate $\cal B$ via the correlation functions $C(a,b)$
through measurements of the joint probabilities $P_{kl}(a,b)$. For
simplicity, we will describe only the recording of $P_{11}(a,b)$ in
detail. Although the field-detection exercise takes place in a pair of
two-dimensional state spaces, in common with Bell tests using particle
detection, a new challenge is presented by the angles $b$ and $b'$ in
stochastic function space. This is because there is no standard technology
to control rotations in an infinite-dimensional function space, and such
control is needed to obtain the required four independent evaluations of
correlation.

\begin{figure}
\begin{center}
\includegraphics[width=6.5cm]{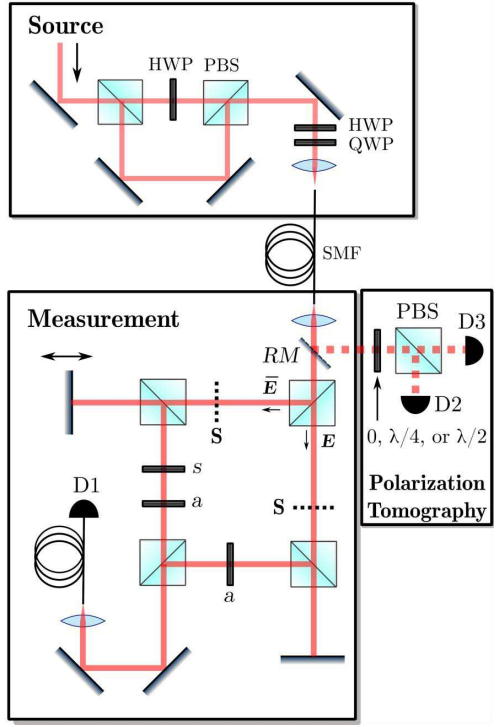}
\caption{\footnotesize The experimental setup consists of a
source of unpolarized light and a measurement using a modified
Mach-Zehnder interferometer.  Half and quarter wave plates (HWP, QWP)
control the polarization of the source.  All beam splitters are 50:50
unless marked as a polarizing beam splitter (PBS).  Intensities
needed for obtaining the required joint probabilities are measured at
detector D1.  Shutters S independently block arms of the
interferometer in order to measure light through the arms separately.
A removable mirror (RM) directs the light to a polarization
tomography setup, where the orthogonal components of the polarization
in the basis determined by the wave plate are measured at detectors
D2 and D3.}
\label{experiment}
\end{center}
\end{figure}

The experimental setup, shown in Figure \ref{experiment}, has two
major components: a source of the light to be measured, and a
Mach-Zehnder (MZ) interferometer.
The source utilizes a 780 nm laser diode, operated in the multi-mode
region below threshold, giving it a short coherence length on the
order of 1 mm. The beam is incident on a 50:50 beam
splitter and recombined on a polarizing beam splitter after adequate
delay so that the light is an incoherent mix of horizontal and
vertical polarizations before being sent to the measurement area via
a single mode fiber.  A half wave plate in one arm controls the
relative power, and thus the degree of polarization (DOP). Quarter and half
wave plates help correct for polarization changes introduced by the
fiber. Polarization tomography is used to characterize the test beam.
Stokes parameters $S_1,\ S_2,\ S_3$ (normalized to $S_0 = 1$) for our
nearly unpolarized source are evaluated as ($ -0.0827,\ -0.0920,\ -0.0158
$), providing DOP = 0.125 (see \cite{Brosseau-Wolf, SupplMat}). This fixes
${\cal B} = 2.817$ as the maximum ideally possible value able to be
achieved for the experimental field.

In Fig.~\ref{experiment} the partially polarized beam entering the MZ is
separated by a 50:50 beam splitter into primary test beam $|{\bf
E}\rangle$ and auxiliary beam $|\bar{\bf E}\rangle$. The two beams
inherit the same statistical properties from their mother beam and
thus both can be expressed as in Eq. (\ref{e-def}), with
corresponding intensities $I$ and $\bar{I}$. The auxiliary beam
obtains an unimportant $i$ phase from the beam splitter.\\

To determine the joint probability $P_{11}(a,b)$ of the test beam
$|{\bf E}\rangle$, the first step is to project the field in the lab
space to obtain $|{\bf E}_1^{a}\rangle \equiv
|{u}^{a}_1\ra\la{u}^{a}_1|{\bf E}\ra$. This can be realized by the
polarizer labelled ${a}$ on the bottom arm of the MZ. The transmitted
beam retains both $|f\ra$ components in function space:
\beq
|{\bf E}_{1}^{a}\ra =\sqrt{I_{1}^{a}}|u _{1}^{a}\ra (c_{11}|f_{1}^{b}\ra +
c_{12}|f_{2}^{b}\ra),
\eeq
where $I_{1}^{a}$ is the intensity, and $c_{11}$ and $c_{12}$ are
normalized amplitude coefficients with $|c_{11}|^{2}+|c_{12}|^{2} = 1$.
Here $c_{11}$ relates to the joint probability in an obvious way:
$P_{11}(a,b) = I_{1}^{a}|c_{11}|^{2}/I$. One sees that the intensities $I$
and $I_{1}^{a}$ can be measured directly but not the coefficient $c_{11}$.

For $P_{11}(a,b)$ our aim is to produce a field that combines a projection
onto  $|f_1^b\ra$ in function-space with the $|u_1^{a}\ra$ projection in
lab space.  The challenge of overcoming the lack of ``polarizers" for
projection of a non-deterministic field onto an arbitrary direction in its
infinite-dimensional function space is managed as follows  by employing
the auxilliary ${\bf \bar E}$ field in the left arm \cite{Qian-Eberly}. We
pass it through the lab space polarizer rotated from the initial basis
$|u_{1}\ra$ by a specially chosen angle $s$, so that the statistical
component $|f_{2}^{b}\ra$ is stripped off. The transmitted beam $|{\bf
\bar E}_{1}^s\ra$ then has only the $|f_{1}^{b}\ra$ component, as desired:
$|{\bf \bar E}_{1}^s\ra = i\sqrt{ \bar{I}_{1}^{s}}|{u}_{1}^{s}\ra
|f_{1}^{b}\ra$. Here $\bar{I}_{1}^{s}$ is the corresponding intensity and
the special angle $s$ is given by \cite{SupplMat,   Qian-Eberly} $\tan s =
(\kappa_{1}/\kappa_{2})\tan b$.

The function-space-oriented beam $|{\bf \bar E}_{1}^s\ra$ is then sent
through another polarizer ${a}$ to become $|{\bf \bar E}_{1}^{a}\ra
=|{u}_{1}^{a}\ra \la {u}_{1}^{a}|{\bf \bar E}_{1}^{a}\ra =
i\sqrt{\bar{I}_1^{a}}|{u}_{1}^{a}\ra  |f_{1}^{b}\ra$, where
$\bar{I}_{1}^{a}$ is the corresponding intensity. Finally, the beams
$|{\bf E}_{1}^{a}\ra$ and $|{\bf \bar E}_{1}^{a}\ra$ are combined by a
50:50 beam splitter which yields the outcome beam $|{\bf E}_{1}^{T}\ra =
(|\bar{\bf E}_{1}^{a}\ra + i|{\bf E}_{1}^{a}\ra)/\sqrt{2}$. The total
intensity $I_{1}^{T}$ of this outcome beam can be easily expressed in
terms of the coefficient $c_{11}$.

Some simple arithmetic will immediately provide the joint probability
$P_{11}(a,b)$ in
terms of various measurable intensities:
\beq \label{eq:prob}
P_{11}(a,b) =(2I_{1}^{T} - \bar{I}_{1}^{a} -
I_{1}^{a})^{2}/4I\bar{I}_{1}^{a}.
\eeq
Other $P_{kl}(a,b)$ values can be obtained similarly by rotations of
polarizers $a$ and $s$. To make measurements, polarizers $a$ are
simultaneously rotated using motorized mounts, while the third polarizer
$s$ is fixed at some value.


For each angle, measurements are made at detector D1 for the total
intensity $I^T$, and the separate intensities from each arm $I^{a}$ and
$\bar{I}^{a}$ by using the shutters $S$ alternately. From these
measurements C(a,b) can be determined and Eq \eqref{CHSH1} is used to
evaluate the
CHSH parameter $\cal B$.

\begin{figure}[t!]
\begin{center}
\includegraphics[width=9cm]{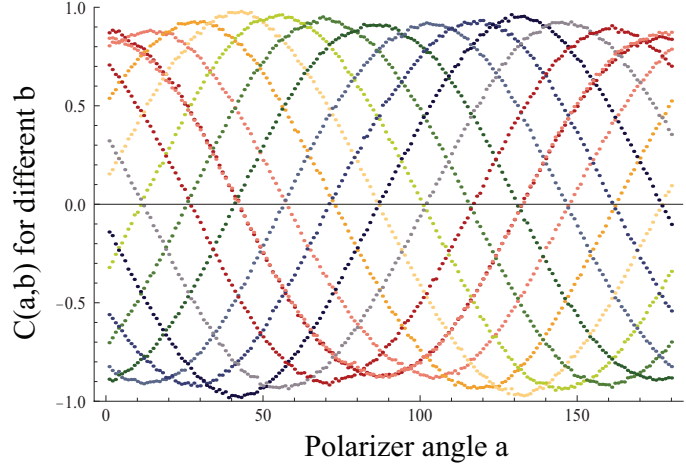}
\caption{\footnotesize Plots of the correlation functions C(a,b) obtained
by rotating
polarizer $a$ in the lab space and holding angle $b$ in the function space
constant. Different curves correspond to different fixed values of $b$
separated by $\pi/12$. The invariant
cosine function required to violate the Bell inequality is clearly
present. Error bars are included but scarcely visible.}
\label{fig:results}
\end{center}
\end{figure}

Fig.~\ref{fig:results} shows $C(a,b)$ obtained by measuring the  joint
probabilities $P_{jk}(a,b)$ for a complete rotation of polarizer $a$, with
different curves corresponding to $b$ (and thus $s$) fixed at different
values. It is apparent from the near-identity of the curves that, to good
approximation, the correlations are a function of the difference in
angles, i.e. $C(a,b)=C(a-b)$. Then the maximum value for
$\cal B$ can be found straightforwardly from any one of the curves in Fig.
~\ref{fig:results}. Among them the smallest and largest values of $\cal B$ are $2.548 \pm 0.004$
and $2.679 \pm 0.007$.

In summary, we defined a field or state to be classical and therefore not
quantum mechanical in any way, but required it to satisfy several
quantum-like conditions. Its bipartite pure state form demonstrates the
clear entanglement of its independent degrees of freedom. This is in
common with pure two-party quantum systems. It is dynamically
probabilistic, meaning that individual field measurements yield values
that cannot be predicted except in an average sense, another feature
shared with quantum systems but also associated for more than 50 years
with well-understood and well-tested optical coherence theory
\cite{Brosseau-Wolf}. Such so-called Shimony-Wolf states that embody this
combination of features have a range of correlation strengths that is
restricted by the conditions of the CHSH Bell inequality. Experimental
tests showed that the Shimony-Wolf states violate the same inequality
proved for them, attaining Bell-violating levels of correlation similar to
those found in tests of quantum systems \cite{brightBell}.

The explanation for the CHSH inequality violation is not hard to find, but
is important because it makes yet another connection between classical
Shimony-Wolf states and quantum systems. We recall that hidden variables
were allowed by Bell (and in the CHSH derivation) to be present and to act
on the bipartite degrees of freedom, and to induce correlations between
them. But so long as the observation made on one of the parts of a tested
system are independent of observations made on the other part, a Bell
inequality will limit those correlation strengths. But, exactly as for
quantum systems, Shimony-Wolf states have an embedded structural
contingency. This contingency is the entanglement of the two examined
degrees of freedom. In effect, we have provided a way of examining a new
sector of the boundary between quantum and classical. The violation
demonstrated here shows that, in
contrast to a wide understanding,  Bell violation has nothing special to
do with quantum mechanics, but everything to do with entanglement.

\noindent {\bf Acknowledgements}
The authors acknowledge helpful discussions over several years with many
colleagues including A.F. Abouraddy, A. Aspect, C.J. Broadbent, L.
Davidovich,  E. Giacobino, G. Howland, A.N. Jordan, G. Leuchs, P.W.
Milonni, R.J.C. Spreeuw, A.N. Vamivakas, as well as financial support from
NSF PHY-0855701, NSF PHY-1203931, and DARPA DSO Grant No.
W31P4Q-12-1-0015.

\section{Supplemental Material}

\section*{Polarization tomography}

To measure the polarization state, polarization analysis is performed by
inserting a removable mirror RM at the input of the Mach-Zehnder
interferometer to send the light to a polarization tomography setup (see
illustration in Fig. 1 in the Letter). Using a polarizing beam splitter
and half and quarter wave plates to project onto circular and diagonal
bases, the Stokes parameters $S_0,\ S_1,\ S_2,\ S_3$ are measured
(normalized values given in the text) and then used for calculation of the
degree of polarization (DOP) (see \cite{Brosseau-Wolf-SM}), i.e.,
\begin{equation}\label{dop}
DOP = \frac{\sqrt{S_{1}^{2}+S_{2}^{2}+S_{3}^{2}} }{S_{0}} = 0.125.
\end{equation}

This is then used to find the Schmidt coefficients $\kappa_1$ and
$\kappa_2$ of the Shimony-Wolf light field:
\beq
\label{e-def} |{\bf E}\ra =\sqrt{I}\Big( \kappa_{1}|u_1\ra|f_1\ra +
\kappa_{2}|u_2\ra|f_2\ra \Big).
\eeq
According to \cite{Qian-EberlyOL-SM} one finds
\begin{equation}\label{kappa-dop}
\kappa_{1},\ \ka_{2} = \sqrt{\frac{1 \pm DOP}{2}} = 0.750,\ \ 0.661.
\end{equation}

\section*{Determining the stripping angle $s$}

This section introduces in detail how a projection in the lab space works
effectively as a stripping of a basis (e.g., the component
$|f_{2}^{b}\rangle $) in function space. The light beam (\ref{e-def}) can
always be rewritten in the rotated function space basis
$|f_{1}^{b}\rangle$, $|f_{2}^{b}\rangle $, i.e,
\begin{eqnarray}
|\mathbf{E}\rangle
&=&\sqrt{I}\Big[(\kappa _{1}\cos b|u_{1}\rangle -\kappa _{2}\sin
b|u_{2}\rangle
|f_{1}^{b}\rangle \notag \\
&&+(\kappa _{1}\sin b|u_{1}\rangle +\kappa _{2}\cos
b|u_{2}\rangle )|f_{2}^{b}\rangle \Big].  \label{before strip}
\end{eqnarray}
Obviously, one notes from the second term of the equation that a properly
chosen polarizer that blocks completely the polarization component $\kappa
_{1}\sin b|u_{1}\rangle +\kappa _{2}\cos b|u_{2}\rangle $ will effectively
strip off the function space basis $|f_{2}^{b}\rangle$.

Such a stripping polarizer can be defined as a rotation of the lab space
basis $|{u}_{1}\rangle$ by an angle $s$, i.e.,
\begin{equation}
|{u}_{1}^{s}\rangle =\cos s|{u}_{1}\rangle -\sin s|{u}_{2}\rangle.
\end{equation}
Then the stripping condition is directly given as
\begin{equation}
0 =\langle {u}_{1}^{s}|(\kappa _{1}\sin b|u_{1}\rangle +\kappa _{2}\cos
b|u_{2}\rangle).
\end{equation}
Some simple arithmetic leads to the following restriction on the rotation
angle $s$:
\begin{equation}
\tan s=(\kappa _{1}/\kappa _{2})\tan b, \label{stripping condition}
\end{equation}
which is determined by the values of $\ka_{1}$ and $\ka_2$ for any
rotation angle $b$.

As a result of this stripping polarizer, the beam (\ref{before strip})
becomes
\begin{eqnarray}
&&|\mathbf{E}_{1}^{s}\rangle =|{u}_{1}^{s}\rangle \langle {u}_{1}^{s}|
\mathbf{E}\rangle \notag \\
&=&\sqrt{I}(\kappa _{1}\cos b\cos s+\kappa _{2}\sin b\sin s)|{u}
_{1}^{s}\rangle |{f}_{1}^{b}\rangle. \label{after strip}
\end{eqnarray}
Apparently, the function space component $|f_{2}^{b}\rangle$ of the
transmitted beam is completely stripped off. By comparing the first term
of Eq.~(\ref{before strip}) with Eq.~(\ref{after strip}), one notes that
this stripping technique passes only a fraction of the light that has the
$|f_{1}^{b}\ra$ component in the function space. Therefore it cannot be
immediately regarded as a projection in the function space.

\section*{CHSH inequality}

In this section we provide details of the derivation of the CHSH
inequality \cite{CHSH-SM} for classical statistical light beams. As
introduced in the Letter, a classical light field can always be
decomposed into the optimum Schmidt form (\ref{e-def}).  To follow
convention we name the lab space containing $|{u}_{1}\ra$, $|{u}_{2}\ra$
as party ``A", and the statistical function space with $|{f}_{1}\ra$,
$|{f}_{2}\ra$ as party ``B".

To examine the beam (\ref{e-def}) in lab space, one makes measurements in
an arbitrary polarization-rotated basis $|{u}_{1}^{a}\ra$,
$|{u}_{2}^{a}\ra$. We designate the measurement result in this basis as
$A(a)$, which takes the maximum value $1$ if all the light is registered
in basis $|{u}_{1}^{a}\ra$, and $-1$ when all the light registered in
$|{u}_{2}^{a}\ra$. Consequently, for the most general case when the light
contains both polarization components the measurement result can be
defined as
\beq
\label{A(a)}A(a)= P({u}_{1}^{a})-P({u}_{2}^{a}),
\eeq
where $P({u}_{k}^{a})$ with $k=1,2$ is the probability of finding the
statistical light beam in lab basis $|{u}_{k}^{a}\ra$. The normalization
condition $P({u}_{1}^{a}) + P({u}_{2}^{a}) = 1$ is satisfied. By this
definition one notes that $A(a)$ is exactly the Stokes parameter $S_1$
\cite{Brosseau-Wolf-SM} in the corresponding basis, and it is continuously
varying between $-1$ and $1$.

Similarly one can characterize the measurement of the light beam in the
effective two-dimensional function space by defining an analogous
measurement result as
\beq
\label{B(b)} B(b)= P({f}_{1}^{b})- P({f}_{2}^{b}),
\eeq
where $P({f}_{l}^{b})$ with $l=1,2$ is the probability of finding the
statistical light beam in function basis $|{f}_{l}^{b}\ra$, and
$P({f}_{1}^{b})+P({f}_{2}^{b})=1$. Note that $B(b)$ is also bounded
between $-1$ and $1$.

In general, an average measurement outcome of one space may condition on
the status of the other space. Moreover, from the argument of Bell
\cite{Bell64-66-SM}, it could also be influenced by some unknown and/or
unmentioned variables and parameters such as environmental noises, the
detector conditions, or more fundamental hidden variables, etc. We follow
tradition in the discussion of Bell inequality monitoring and label these
so-called ``contextual" classical unknowns collectively by a single
multi-dimensional parameter $\{\lambda\}$ with an overall unknown
distribution $\rho(\{\lambda\})$. Therefore the measurement result $A(a)$
in lab space can be rewritten by admitting all such dependences:
$$A(a) \to A(a,\{\lambda\}|B),$$
and similarly
$$B(b) \to B(b,\{\lambda\}|A).$$
Then the measurement outcome correlation between the lab and function
spaces can be characterized as
\beq \label{CAB1}
C(a,b) = \int A(a,\{\lambda\}|B )B(b,\{\lambda\}|A )\rho
(\{\lambda\})d\{\lambda\}. \eeq

As usual, measurement outcomes in ``A" space are assumed to be independent
of measurements and setups made in ``B" space, and vice versa, so we have
in terms of measurement results the simpler forms
$A(a,\{\lambda\}|B)=A(a,\{\lambda\})$ and
$B(b,\{\lambda\}|A)=B(b,\{\lambda\})$. It is important to note that this
assumption does not exclude possible correlations between the measurements
in the two spaces, i.e., the outcomes in both spaces may still be related
because of $\{\lambda\}$. Consequently the correlation function
(\ref{CAB1}), for arbitrary angles $a$ and $b$, is equivalent to
\beq \label{CAB2}
C(a,b) = \int A(a,\{\lambda\} )B(b,\{\lambda\} )\rho
(\{\lambda\})d\{\lambda\}.
\eeq

Then one can follow the standard CHSH procedure \cite{CHSH} and find
\begin{eqnarray}
\label{CHSH}
\mathcal{B} &=&C(a,b)-C(a,b')+C(a',b)+C(a',b') \notag \\
&=&\int \rho(\{\lambda \})d\{\lambda \} \Big[A(a,\{\lambda
\})\Big(B(b,\{\lambda \})-B(b',\{\lambda \})\Big) \notag \\
&&+A(a',\{\lambda \})\Big(B(b,\{\lambda \})+B(b',\{\lambda \})\Big)\Big].
\end{eqnarray}
From the fact that any measurement results $A(a,\{\lambda \})$ and
$B(b,\{\lambda \})$ lie between the values $-1$ and $1$, the expression
for $\mathcal{B}$ straightforwardly obeys the familiar inequality $-2\leq
\mathcal{B}\leq 2$ for any $a$, $a'$, $b$, $ b'$. This concludes the
derivation of the CHSH inequality.


\begin{thebibliography}{99}

\bibitem{Simon-etal} B.N. Simon, S. Simon, F. Gori, M. Santarsiero, R.
Borghi, N. Mukunda, and R. Simon, ``Nonquantum Entanglement Resolves a
Basic Issue in Polarization Optics", \prl{104}, 023901 (2010). See also M.
Sanjay Kumar and R. Simon, ``Characterization of Mueller matrices in
polarization optics", \oc{88}, 464 (1992), and F. T\"oppel, A.
Aiello, C. Marquardt, E. Giacobino, and G. Leuchs, ``Classical
entanglement in polarization metrology", arXiv:1401:1543 (2014).

\bibitem{Qian-EberlyOL} X.F. Qian and J.H. Eberly, ``Entanglement and
classical polarization states", \ol{36}, 4110 (2011). See also T.
Set\"al\"a, A. Shevchenko and A. Friberg, ``Degree of polarization for
optical near fields", \pre{66}, 016615 (2002), and J. Ellis and A.
Dogariu, ``Optical polarimetry of random fields", \prl{95}, 203905 (2005).

\bibitem{Kagalwala-etal} K.H. Kagalwala, G. DiGiuseppe, A.F. Abouraddy and
B.E.A. Saleh, ``Bell's measure in classical optical coherence", \npho{7},
72 (2013).

\bibitem{Spreeuw} R.J.C. Spreeuw, ``A Classical Analogy of Entanglement",
Found. Phys. {\bf 28}, 361-374 (1998).

\bibitem{Ghose-Samal01} P. Ghose and M.K. Samal, ``EPR Type
Nonlocality in Classical Electrodynamics",
arXiv:quant-ph/0111119 (2001).

\bibitem{Lee-ThomasPRL} K.F. Lee and J.E. Thomas, ``Entanglement with
classical fields" \prl {88}, 097902 (2002).

\bibitem{Borges-etal} C.V.S. Borges, M. Hor-Meyll, J.A.O. Huguenin, and
A.Z. Khoury, ``Bell-like Inequality for the spin-orbit separability of a
laser beam", \pra{82}, 033833 (2010).

\bibitem{Goldin-etal} M.A. Goldin, D. Francisco and S. Ledesma,
``Simulating Bell inequality violations with classical optics encoded
qubits", \josa{27}, 779 (2010).

\bibitem{Shimony} A. Shimony, ``Contextual Hidden Variables Theories and
Bell's Inequalities", Brit. J. Phil. Sci. {\bf 35}, 25-45 (1984). See also
``Bell's Theorem", A. Shimony, in {\em Stanford Encycl. of Philosophy},
http:/plato.stanford.edu/entries/bell-theorem  (2009).

\bibitem{Wolf59} E. Wolf, ``Coherence Properties of Partially Polarized
Electromagnetic Radiation", N. Cim. {\bf 13}, 165 (1959).

\bibitem{Freedman-Clauser} S.J. Freedman and J.F. Clauser, ``Experimental
Test of Local Hidden-Variable Theories", \prl{28}, 938 (1972).

\bibitem{Aspect-etal} A. Aspect, P. Grangier and G. Roger, ``Experimental
Realization of Einstein-Podolsky-Rosen-Bohm Gednakenexperiment: a new
violation of Bell's Inequalities", \prl{49}, 91 (1982); and A. Aspect, J.
Dalibard, and G. Roger, ``Experimental Test of Bell's Inequalities Using
Time-Varying Analyzers", \prl{49}, 1804 (1982).

\bibitem{Ou-Mandel} Z.Y. Ou and L. Mandel, ``Violation of Bells inequality
and classical probability in a 2-photon correlation experiment", \prl{61},
50 (1988).

\bibitem{Shih-Alley} Y.-H. Shih and C.O. Alley, ``New Type of
Einstein-Podolsky-Rosen-Bohm Experiment Using Pairs of Light Quanta
Produced by Optical Parametric Down Conversion", \prl{61}, 2921 (1988).

\bibitem{Rowe-etal} M.A. Rowe, D. Kielpinski, V. Meyer, C.A. Sackett, W.M.
Itano, C. Monroe, and D.J. Wineland, ``Experimental violation of a Bell's
inequality with efficient detection", \nat{409}, 791 (2001).

\bibitem{Schmidt} See, for example, A. Ekert and P.L. Knight, ``Entangled
Quantum-Systems and the Schmidt Decomposition", \ajp{63}, 415 (1995). The
original paper is: E. Schmidt, ``Zur Theorie der linearen und
nichtlinearen Integralgleichungen. 1. Entwicklung willk\"uriger Funktionen
nach Systeme vorgeschriebener", Math. Ann. {\bf 63}, 433 (1907). The
Schmidt theorem is a continuous-space version of the singular-value
decomposition theorem for matrices. For background, see M.V. Fedorov and
N.I. Miklin, ``Schmidt modes and entanglement", Contem. Phys. {\bf 55}, 94
(2014).

\bibitem{Kac-Siegert} The functions can be obtained as eigenfunctions
of integral equations in which the kernel is a component's individual
autocorrelation function. See M. Kac and A.J.F. Siegert,  ``An
explicit representation of a stationary gaussian process", Ann. Math.
Stat. {\bf 18} 438-442 (1947), and L. Mandel and E. Wolf, ``Optical
Coherence and Quantum Optics", Cambridge Univ. Press (1995), Chap. 2.

\bibitem{Bell64-66} J.S. Bell, ``On the Einstein Podolsky Rosen Paradox",
Physics {\bf 1}, 195 (1964) and J.S. Bell, ``On the problem of hidden
variables in quantum mechanics", \rmp{38}, 447-452 (1966).

\bibitem{BellSpeakable} J.S. Bell, {\em Speakable and Unspeakable in
Quantum Mechanics} (Cambridge Univ. Press, Cambridge, 2004), 2nd edition.
Essays 4, 7, 16, and 24 are particularly relevant here.

\bibitem{CHSH} J.F. Clauser, M.A. Horne, A. Shimony and R.A. Holt,
``Proposed Experiment to Test Local Hidden-Variable Theories", \prl{23},
880 (1969).

\bibitem{Brosseau-Wolf} See C. Brosseau, {\em Fundamentals of Polarized
Light: A Statistical Optics Approach} (Wiley, New York, 1998) and  E.
Wolf, {\em Introduction to the Theory of Coherence and Polarization of
Light} (Cambridge Univ. Press, 2007)

\bibitem{Gisin} N. Gisin, ``Bell's inequality holds for all non-product
states", \pla{154}, 201 (1991).

\bibitem{SupplMat} See supplemental material.

\bibitem{Qian-Eberly} See X.-F. Qian and J.H. Eberly, ``Entanglement is
Sometimes Enough",
arXiv:1307.3772.

\bibitem{brightBell} See Paul G. Kwiat, Klaus Mattle, Harald Weinfurter,
Anton Zeilinger, Alexander V. Sergienko, and Yanhua Shih, ``New
High-Intensity Source of Polarization-Entangled Photon Pairs", \prl{75},
4337 (1995), and Paul G. Kwiat, Edo Waks, Andrew G. White, Ian Appelbaum,
and Philippe H.
Eberhard, ``Ultrabright source of polarization-entangled photons",
\pra{60}, R773(R) (1999).


\end{thebibliography}

\begin{thebibliography}{99}

\bibitem{Brosseau-Wolf-SM} See C. Brosseau, {\em Fundamentals of Polarized
Light: A
Statistical Optics Approach} (Wiley, New York, 1998) and  E. Wolf, {\em
Introduction to the Theory of Coherence and Polarization of Light}
(Cambridge Univ. Press, 2007).

\bibitem{Qian-EberlyOL-SM} See X.F. Qian and J.H. Eberly, ``Entanglement and
classical polarization states", \ol{36}, 4110 (2011).

\bibitem{CHSH-SM} J.F. Clauser, M.A. Horne, A. Shimony and R.A. Holt,
``Proposed Experiment to Test Local Hidden-Variable Theories", \prl{23},
880 (1969).

\bibitem{Bell64-66-SM} J.S. Bell, ``On the Einstein Podolsky Rosen Paradox",
Physics {\bf 1}, 195 (1964) and J.S. Bell, ``On the problem of hidden
variables in quantum mechanics", \rmp{38}, 447-452 (1966).

\end{thebibliography}
\end{document}